# Experimental Test of Decoherence Theory using Electron Matter Waves

Peter J. Beierle,[1] Liyun Zhang,[2] and Herman Batelaan[1]*

[1]Department of Physics and Astronomy, University of Nebraska-Lincoln, Lincoln Nebraska 68588, USA

[2] Key Laboratory of Quantum Information and Quantum Optoelectronic Devices, Xi'an Jiaotong University, Xi'an 710049, People's Republic of China

A controlled decoherence environment is studied experimentally by free electron interaction with semi-conducting and metallic plates. The results are compared with physical models based on decoherence theory to investigate the quantum-classical transition. The experiment is consistent with decoherence theory and rules out established Coulomb interaction models in favor of plasmonic excitation models. In contrast to previous decoherence experiments the present experiment is sensitive to the onset of decoherence.

The continuous divide between quantum and classical physics can be described by decoherence theory. Decoherence is an irreversible process in which a quantum state entangles with an environment in such a way that it loses its interference properties [1,2]. For most experiments, maintaining a system's quantum coherence is desirable, and great efforts are made to isolate the system from its environment [3–6]. Additionally, it has been suggested that some sources of decoherence may be ubiquitous, such as those originating from vacuum field fluctuations or gravitation [7–12], and that decoherence in general is a critical element in resolving the quantum measurement problem [13]. Thus, experimentally sorting out various sources of decoherence and determining which dominate is desirable for both technical applications and fundamental studies, including the decoherence program [13].

There have been experiments in which the transition between the quantum and classical domain has been controlled through both the "distance" between states [14–16] and the strength of the interaction with the environment [16–20]. Most of these experiments involve various matter-wave interferometric techniques.

In this Letter, we will describe a decoherence setup that is a realization of Zurek's original thought experiment of diffracting charges through a grating and controlling the spatial quantum coherence with a conducting surface [21]. We have measured the effect of a gold and silicon surface and found upper bounds on the loss of contrast due to decoherence. These results refute current decoherence models premised on image charge [22–24]. We also identify viable decoherence models based on dielectric excitation theory from effects including surface plasmons [25,26].

Sonnentag and Hasselbach previously used an electron biprism interferometer setup with separated arms passing over a semi-conducting surface before recombination [16]. In contrast, we used electron diffraction from a nano-grating and measured the effect of a conducting surface. As we will show below, diffraction is well suited for measuring small losses in coherence, which is particularly useful to detect weak decoherence channels. Sonnentag's and Hasselbach's measurements on doped n-type silicon reveals a decoherence strength that is a factor of $\approx 10^2$ too weak as compared to Zurek's image charge model. This is confirmed by our findings.

The determination of the physical mechanism nevertheless supported image charge models [16,23] as the analysis ignored the strength of decoherence and was limited to a best fit of the functional form, as was done in a similar experiment by Röder and Lubk [27]. The implicit assumption is that a metallic surface (as used in the theory) behaves similarly as a silicon surface. The image charge models were thus considered valid. Our measurements, which now also includes the conductor gold as well as silicon, refutes this conclusion and identifies Howie's model [25,26] as viable.

A 1.67 keV electron beam (Kimball EGG-3101 electron gun) is sent through two collimation slits separated by 25 cm with a geometrical beam divergence of 61 μrad in the x-direction and 120 μrad in the y-direction (see Fig. 1). This collimation gives a transverse coherence length of the electron beam of approximately 600 nm as determined by diffraction images. This makes it possible to diffract the electrons from a 100 nm periodic nanofabricated grating [28,29]. The diffracted electron distribution is magnified 24 cm downstream by an electrostatic quadrupole lens, detected by a multichannel plate detector, backed by a phosphorous screen (Beam Imaging Solutions BOS-18), and imaged by a CCD camera. A LabVIEW image acquisition program [30] accumulates a two-dimensional streaming image from the camera. The vacuum chamber in which this experiment takes place is held at a pressure of $\approx 4 \times 10^{-5}$ Pa and is protected from external magnetic fields by two layers of mu-metal magnetic shielding.

A 1 cm² surface is then brought in from below the diffracted beam 3 mm after the grating such that the surface in the $x$-$y$ plane is perpendicular to the diffraction grating bars. The surface height is adjusted to cut into the beam so that 1/3 of the intensity of the original beam reaches the detector. The surface is supported by a mechanical feedthrough whose angular pitch with respect to the beamline can be adjusted with a precision of approx. 0.2 mrad. This pitch of the surface is adjusted to maximize the electron beam's deflection due to image charge attraction. The Si surface was cleaned using a version of the industry-standard RCA cleaning method (without the oxide strip) [31], to remove dust or other contaminants.

When an electron passes over a decohering surface, it interacts with the surface so that the interference pattern in the far field has lower visibility, and further decreases the closer the electron passes over the surface. Previously, the decoherence was measured in terms of the visibility of the interference pattern [14,18,16,27], i.e. $V_{is} = (I_{max} - I_{min})/(I_{max} + I_{min})$.

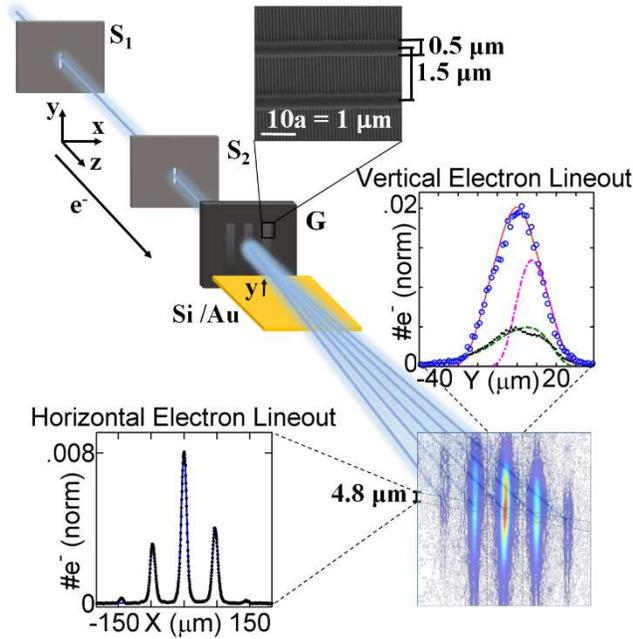

FIG. 1. Experimental Setup. Electrons are prepared in a spatially coherent state by collimation with two slits (S$_1$ and S$_2$), then diffracted through a nanofabricated diffraction grating (G) before passing over either a doped Silicon (Si) or Gold (Au) surface, which acts as the decohering "environment". Middle-Right: Electron distribution in the $y$-direction taken when there is no wall (solid red line) compared to when the surface is raised to cut 1/3 of the beam (dashed and dotted-dashed lines). The distribution in the $y$-direction closely fits the simulation when image charge is present (dashed line) as opposed to no image charge (dashed dotted line).

However, in order to measure smaller changes in contrast and reduce the uncertainty in measurement due to background counts, we measure coherence in terms of the transverse coherence length of the diffracted beam as observed at the detector [32]:

$$L_{coh} \approx \frac{\lambda_{dB}}{\theta_{coll}} \approx \frac{ad}{w_{FWHM}}. \quad (1)$$

Here $a$ is the periodicity of the grating, $d$ is the distance between neighboring diffraction peaks at the detection screen, and $w_{FWHM}$ is the FWHM of a diffraction peak. Thus, here we associate a loss of coherence with an increase of the width of the diffraction peaks $w_{FWHM}$ rather than a loss of visibility.

Two dimensional images of the electron interference pattern are recorded. Line-outs of the images are extracted to obtain diffraction patterns. The line-outs are taken at a slant with the $x$-direction to compensate for image skew. In the $y$-direction a 4.8 μm range on the detector is integrated for each line-out at vertical position Y. The diffraction line-outs are modeled by,

$$I(x) = \left\{\frac{\sin[\alpha(x-x_0)]}{\alpha(x-x_0)}\right\}^2 \left[\sum_{n=0} G(x-x_1 \pm nd)\right] + A_{bckd} \exp\left[-(x-x_2)^2/2c_3^2\right], \quad (2)$$

where the first term corresponds to the single slit envelope and the diffraction peaks. Each individual peak

$$G(x) = a_1 e^{-x^2/2c_1^2} + (1-a_1) e^{-x^2/2c_2^2}, \quad (3)$$

is approximated by two Gaussians with overlapping means, which also very well fits the shape of the beam without diffraction. The width of all diffraction peaks is constrained to be the constant, as is the peak to peak distance $d$ is taken to be constant. From this fit the $w_{FWHM}$ and $d$ is extracted to compute the coherence length $L_{coh}$ for a given distribution according to Equation (1).

The advantage of using diffractometry over interferometry lies in their respective decoherence measures, $L_{coh}$ and $V_{is}$. The background signal can be subtracted for diffraction without distorting the measured value of $L_{coh}$. This is not the case when measuring visibility in an interferometer. The visibility $V_{is}$ drops off linearly due to a weak background signal, which can mask decoherence. For a weak decohering environment that scatters the incident beam and introduces background, diffractometry is thus well suited.

There are various models that predict different coherence lengths as a function of height above the surface, and material properties of the surface. The original model that focused on electron-surface decoherence was conceived of by Anglin and Zurek [21,22]. The physical system is a classical image charge on the surface of the conductor that follows the free electron as it travels parallel to the surface. Joule heating, experienced by the image charge while traversing the surface, causes dissipation with a relaxation time $\tau_{relax}$. Back-action on the free electron leads to decoherence with a corresponding time $\tau_{dec}$ (called decorrelation time in [33]). The decoherence time is taken to be proportional to the relaxation time according to [33],

$$\tau_{dec} = \left(\frac{\lambda_{th}}{\Delta x}\right)^2 \tau_{relax}, \quad (4)$$

Where $\lambda_{th}$ is the thermal de Broglie wavelength of the image charge, and $\Delta x$ is the distance between decohering paths associated with the electron moving over the surface. the relaxation time due to Ohmic dissipation is $\tau_{relax} = 16\pi m y^3 / e^2 \pi \rho$. Thus, according to this model the resulting decoherence time scale is [34],

$$\tau_{dec}^{Zurek} = \frac{4h^2}{\pi e^2 k_B T \rho} \frac{y^3}{(\Delta x)^2}, \quad (5)$$

where $\rho$ corresponds to the resistivity of the surface material and $e$ is the electron's charge, $y$ is the height of the electron over the surface, $m$ is the mass of the electron, $k_b$ is Boltzmann's constant, and $T$ is temperature of the surface (room temperature).

The decoherence model by Scheel and Buhmann [24] is also based on the electron's interaction with its image charge, but it considers a full "macroscopic quantum electrodynamics" treatment. This takes into consideration the surface's linear dielectric response. Taking the low frequency limit where the Drude approximation $\varepsilon(\omega) \approx 1 + i/(\varepsilon_0 \rho \omega)$ holds for both gold [24] and doped silicon [25,35], the decoherence time scale is,

$$\tau_{dec}^{Buhmann} = \frac{\pi \varepsilon_0 \hbar^2}{e^2 k_B T \rho}\left[\frac{1}{2y} - \frac{1}{\sqrt{(2y)^2 - (\Delta x)^2}}\right]^{-1}. \quad (6)$$

In the limit $\Delta x \ll y$ this is equivalent to Equation 5.

While these aforementioned models posit to hold for a wide range of materials (including semiconductors and metals) Howie's model [25] is worked out only for semiconductors and Machnikowski's model [23] is only worked out for metals. Howie's model is based on event probability $e^{-P}$ rather than energy dissipation, where such events correspond to aloof scattering with long wavelength plasmons and "similar excitations" up to a cutoff frequency $\omega_m = 0.6 \times 10^{12}$ Hz. The stated expression for this probability for a path length $L$ is,

$$P^{Howie} = \left(\frac{e^2 L \omega_m^2}{4\pi^2 \hbar \sigma v^2}\right) \int_{y/4\Delta x}^{\infty} \frac{\exp(-s)}{s} ds, \quad (7)$$

where $L$ is the length of the surface, $\sigma$ is the conductivity of the surface $v$ and is the velocity of the electron in the $z$-direction. The exponential integral is approximated by [36],

$$-Ei(-\eta) = \int_{\eta}^{\infty} \frac{\exp(-s)}{s} ds \cong \left(A^{-7.7} + B\right)^{-0.13}, \quad (8)$$

where $\eta = y/4\Delta x$ [37],

$$A = \log\left[\left(0.56146/\eta + 0.65\right)(1+\eta)\right] \quad \& \quad B = \eta^4 e^{7.7\eta}(2+\eta)^{3.7}.$$

Machnikowski's fully quantum many-body electron gas model infers that the primary decoherence mechanism is due to the dissipative effects of image charge formation rather than Ohmic resistivity effects [23]. It is notably dependent on the Fermi wave-vector for metals $(k_{Fermi})$. This decoherence time scale is,

$$\tau_{dec}^{Machnikowski} = \frac{32\varepsilon_0 \hbar^2 k_{Fermi}}{\pi e^2 m k_B T}\left(\frac{y}{\Delta x}\right)^2. \quad (9)$$

Decoherence over the surface modifies the density matrix of the electron according to [1,38]:

$$\rho_{final} = \rho_{initial} e^{-\int_{t_i}^{t_f} dt/\tau_{dec}} = \rho_{initial} e^{-\Gamma}, \quad (10)$$

where $\Gamma$ is the decoherence factor and the decoherence time scale $\tau_{dec}$ is not only model-dependent, but also depends on $\Delta x$ and $y$. For all of the models, the diffraction pattern is obtained by propagating the final density matrix to the detection screen (see Supplementary material [39] for details). The change in transverse coherence length is then obtained from the calculated far-field diffraction pattern using Equation 1.

Plotted in Fig. 2 is a comparison of the coherence length as a function of height for the case of two different n-type phosphorous doped silicon samples of resistivities 1-20 $\Omega$cm and 1-10 $\Omega$cm (data points). Our results agree with Hasselbach's experimental findings, who used a 1.5 $\Omega$cm n-type doped silicon sample of 1 cm length using the same beam energy of 1.67 keV. The observed loss of contrast can be visualized in the diffractogram's diffraction peak broadening (Fig. 2 top right), based on the histogram data collected from the CCD camera (see Supplementary material [39] for more details).

The experimental data is also compared to Zurek's model of classical image charge/Ohmic dissipation, Scheel and Buhmann's macroscopic quantum electrodynamic model and Howie's dielectric excitation theory model. The uncertainty associated with the theoretical curves (shaded regions I, II and III) in Fig. 2 corresponds to the range of Si resistivity 1-20 $\Omega$cm. The shaded region for Hasselbach's experimental fit (IV) corresponds to the published experimental uncertainty [16].

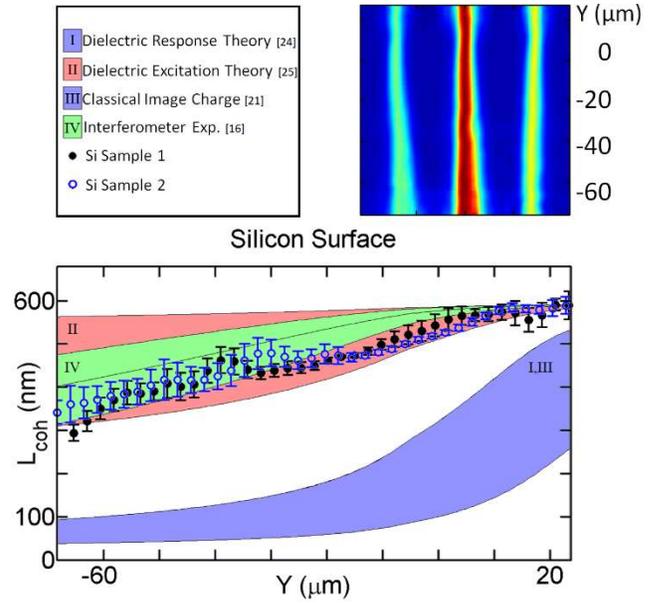

FIG. 2. Transverse Coherence length for a silicon surface. The diffraction pattern shows a loss of contrast as the diffraction peaks broaden for electrons that passed closer to the surface (top right). Our experimental findings (dotted) show agreement with Hasselbach's experimental fit (IV, green), and is consistent with modelling based on dielectric excitation theory (II, red). The data does not agree with models based on Ohmic dissipation due to classical image charge and macroscopic quantum electrodynamic theory using dielectric response (I and III, blue).

The observed loss of contrast in doped-silicon rules out Zurek's and Scheel's decoherence models. This is in contrast to the claim made earlier that Zurek's model is in adequate agreement with experiment [16]. Even if dephasing is present in our experiment, the observed loss of contrast is much smaller than predicted by the models and therefore the conclusion remains valid. Howie's dielectric excitation model is in agreement with our findings.

This experiment was also carried out for the case of a gold surface, and plotted in Fig. 3 is the transverse coherence length as a function of height. For a metal with a resistivity of 2.2x10$^{-6}$ $\Omega$cm [34], no reduction in contrast is measured for an electron passing close to the gold surface. This is consistent with Zurek's and Scheel & Buhman's models. Machnikowski's image charge formation model significantly overestimates the loss of coherence, despite being developed for high conductivity metals such as gold. Hence, Machnikowski's model can also be ruled out as a viable decoherence mechanism.

The general lack of height-dependence of the loss of contrast can be visualized in the diffraction peak's width of the diffraction pattern remaining approximately constant (Fig. 3 top right). This height independence of the coherence length for the case of the gold surface contrasts that of doped silicon. This may be connected to the much smaller resistivity of gold than doped silicon. No theoretical model is currently able to explain both results.

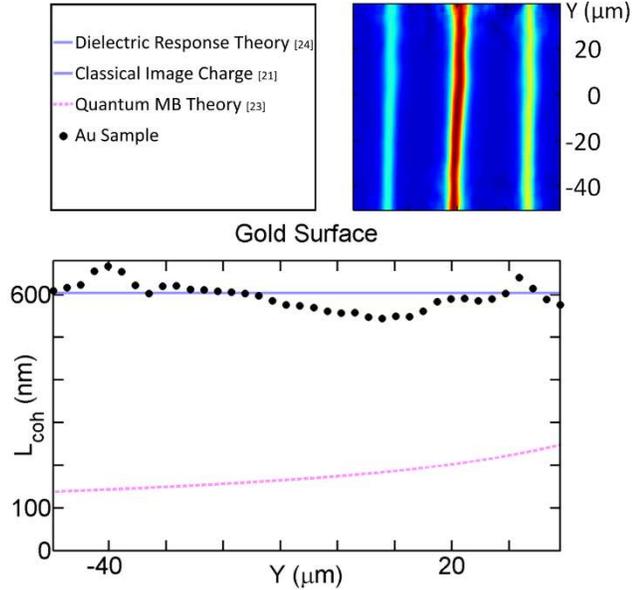

FIG. 3. Transverse Coherence length for a gold surface. The diffraction pattern shows no loss of contrast for electrons that passed close to the surface (top right). The data does not agree with the image charge formation model in a quantum many-body electron gas model (dashed line, pink). The predicted decoherence is too strong and can be ruled out.

This nano-grating diffraction setup opens the door to more sensitive measurements of weak decoherence results. Consider that our modest experimental setup is limited by an initial coherence width ($\approx 600$ nm) and that the decoherence factor in many cases scales as $(\Delta x)^2$. Given that it is now possible for transmission electron microscopes (TEM) to reach coherence lengths as large as 100 μm [41], the sensitivity can thus be improved by $\approx 10^4$. The general method of detection present here opens the pathway to study spatially dependent decoherence surface effects due to plasmon excitation [25,26,42], optical bandgap excitation, superconductive transitions, spin dependent transport effects [43–45], coherent thermal near-fields [46–48], blackbody-like near-fields [49,50], etc.

In conclusion, we have confirmed the loss of contrast in an electron diffraction pattern due to the introduction of a doped silicon surface with a strength consistent with Sonnentag and Hasselbach's biprism interferometer experiment. Our diffractometer setup is simpler in terms of its components and is particularly advantageous in observing weak decoherence effects. Thus, we have shown a new pathway to observe weak decoherence channels. Additionally, for the case of a gold surface we have placed an upper bound on the loss of contrast that can be attributed to decoherence. The silicon and gold decoherence results together confirm that the observed effect is strongly material dependent. We have ruled out a range of decoherence models due image charge based on classical theory [22], quantum many body theory [23], and dielectric theory [24]. For the materials and electron beam parameter range studied, our work remains consistent with decoherence effects due to dielectric excitation theory from effects including surface plasmons [25,26]. These findings are consistent with the general decoherence program [1,2,13].

We thank Vijay Singh, Keith Foreman and Stephen Ducharme for their help in surface preparation. Characterization analysis was performed at the NanoEngineering Research Core Facility (NERCF), University of Nebraska-Lincoln. This work was completed utilizing the Holland Computing Center of the University of Nebraska, which receives support from the Nebraska Research Initiative. We gratefully acknowledge support by the U.S. National Science Foundation under Grant No. 1602755.

# Experimental Test of Decoherence Theory using Electron Matter Waves – Supplemental material

Peter J. Beierle,[1] Liyun Zhang,[2] and Herman Batelaan[1]*

[1]Department of Physics and Astronomy, University of Nebraska-Lincoln, Lincoln, Nebraska 68588, USA
[2]Key Laboratory of Quantum Information and Quantum Optoelectronic Devices, Xi'an Jiaotong University, Xi'an 710049, People's Republic of China

## THEORETICAL METHOD

In order to compute the final coherence length as a function of height Y on the far field detector as predicted by the various physical models, we used a combination of simulating classical trajectories in the *y*-direction and evolution of the electron's density matrix in the *x*-direction as it passes over the surface. The *z*-direction propagation is given by $z = vt$ (see Fig. 1 in [1]). The motivation for separating the motion in a classical and quantum part is that the electron beam collimation slits are tall in the *y*-direction and yield a very small transverse coherence length, justifying a classical approach for motion in the *y*-direction. In the *x*-direction, the slits are narrow and yield a coherence length wider than the grating periodicity, necessitating a quantum approach as intended by the design of the experiment.

By simulating classical trajectories in the *y-z* plane, we can match the distribution of trajectories to the experimentally measured intensity distribution at the detector (Fig. 1). Starting with a distribution of initial positions and momentum defined by the 1st and 2nd slits, the trajectories over the surface are computed in time-steps including an image charge force on the electrons in the *y*-direction. The surface cuts into the electron beam so that only 1/3 of the original electron beam flux makes it to the detector. The electron's free propagation between the surface and the detector is computed and the position on the detector is binned and recorded. This classical simulation well approximates what is observed experimentally at the detector (Fig. 1).

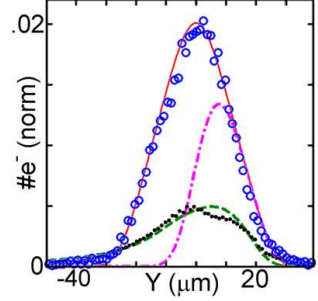

FIG. 1. Electron distribution in the *y*-direction. The case without a wall (solid red line) is compared to when the surface is raised to cut 1/3 of the beam (dashed and dotted-dashed lines). The distribution in the *y*-direction closely fits the simulation when image charge is present (dashed green line) as opposed to when no image charge is present (dashed dotted magenta line).

We first prepare the initial density matrix of the free electron by considering a partially coherent Gaussian beam,

$$\rho(x, x') = \frac{1}{\sigma_0^{coh}\sqrt{2\pi}} \exp\left[-(x-x_0)^2 / 2(\sigma_0^{coh})^2\right] \times \exp\left[-(x')^2 / 2(\sigma'_{initial})^2\right] \quad (1)$$

Here $x$ and $x'$ describe the coordinates of the matrix element in the direction of the diagonal and in the direction orthogonal to the diagonal respectively (Fig. 2). The position $x_0$ indicates the center of the Gaussian. The width of the Gaussian in the $x'$-direction, $w'_{initial} \equiv 2\sqrt{2\ln(2)}\sigma'_{initial}$, is proportional to the transverse coherence length. The spatial width along $x$, $w \equiv 2\sqrt{2\ln(2)}\sigma_0^{coh}$ is determined by a path integral simulation taking into consideration propagation through the first two collimation slits and reaching the beginning of the surface [2]. Note that if $w'_{initial}$ equals $w$, then the initial beam is fully coherent. If $w'_{initial}$ is smaller than $w$, then the initial beam is partially coherent as in Fig. 2 (left). The initial state of the electron $\rho_{initial}$ describes the electron before interacting with the surface at time $t_i$.

Now we consider the electron interaction with the surface. We model the change in transverse coherence length due to the interaction from a given decoherence process by considering the evolution of the density matrix of the electron. It changes according to [3]:

$$\rho_{final} = \rho_{initial} e^{-\int_{t_i}^{t_f} dt/\tau_{dec}}, \quad (2)$$

where the decoherence time scale $\tau_{dec}$ is model-dependent, and depends on $\Delta x$ and $y(t)$. When computing the integral in Equation 2, the simulated trajectories $y(t)$ are inserted.

Each element in the density matrix is computed according to equation 2. The distance between paths $\Delta x$ used in the models equals the distance between the corresponding off diagonal terms in the density matrix (Fig. 2). The final state of the electron $\rho_{final}$ is now found right after the interaction with the surface at time $t_f$. The density matrix of the electron has the form,

$$\rho_{final}(x,x') = \frac{1}{\sigma_0^{coh}\sqrt{2\pi}} \exp\left[-(x-x_0)^2 / 2(\sigma_0^{coh})^2\right] \\ \times \exp\left[-(x')^2 / 2(\sigma'_{final})^2\right], \quad (3)$$

where the width of the final state orthogonal to the diagonal is smaller than the width of the initial state ($w'_{final} < w'_{initial}$). This step describes decoherence.

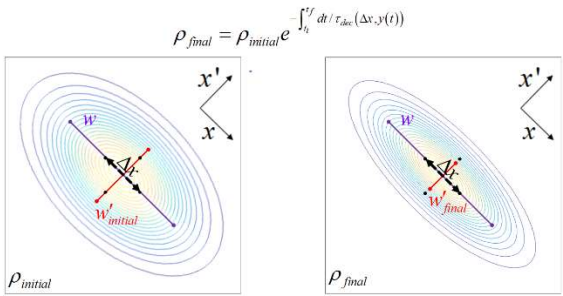

FIG. 2. Evolution of density matrix. As a result of decoherence, the initial state (left) evolves such that the off-diagonal elements reduce in amplitude. Hence the state's width $w'$ decreases (right).

Making use of the ability to write a partial coherent state as a sum of coherent (i.e. pure) states (see Fig. 3),

$$\rho_{final} = \sum_{n=1}^{\infty} c_n \rho_n^{coh}, \quad (4)$$

then we can write $\rho_{final}$ as a sum of Gaussian coherent states,

$$\rho_{final} \approx \sum_{n=1}^{N} \exp\left[-(x-x_n)^2 / 2(\sigma_{env})^2\right] \\ \times \exp\left\{-\left[(x-x_0-x_n)^2 + (x')^2\right] / 2(\sigma_2^{coh})^2\right\}, \quad (5)$$

where $\sigma_2^{coh} = \sigma'_{final}$ describes the width of the reduced pure states after decoherence and

$$\sigma_{env} = \sqrt{(\sigma_0^{coh})^2 - (\sigma_2^{coh})^2}, \quad (6)$$

is the width of the envelope of the convolution. The infinite sum in Eq. 4 is approximated numerically with a finite sum in Eq. 5.

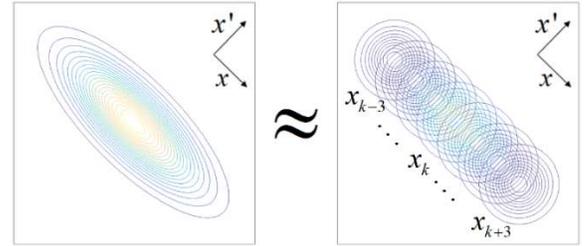

FIG. 3. Deconvolution of a partial coherent (mixed) state by a series of coherent (pure) states.

Next, each wave function corresponding to one of the reduced pure states is acted upon by a grating function. A Fourier transform is used to determine the far field pattern. This is repeated for each of the reduced pure states and the resulting probability distribution patterns are summed to give the final far field diffraction pattern (Fig. 4 bottom right). It is from this final pattern that a transverse coherence width $L_{coh}(Y)$ is computed using $L_{coh} \approx \lambda_{dB}/\theta_{coll} \approx ad/w_{FWHM}$, where $a$ is the periodicity of the grating, $w_{FWHM}$ is the width of the computed diffraction peaks in the far field, and $d$ is the distance between diffraction peaks. It is these values $L_{coh}(Y)$ which produce the theoretical curves in the Fig. 2 and Fig. 3 in the main paper [1].

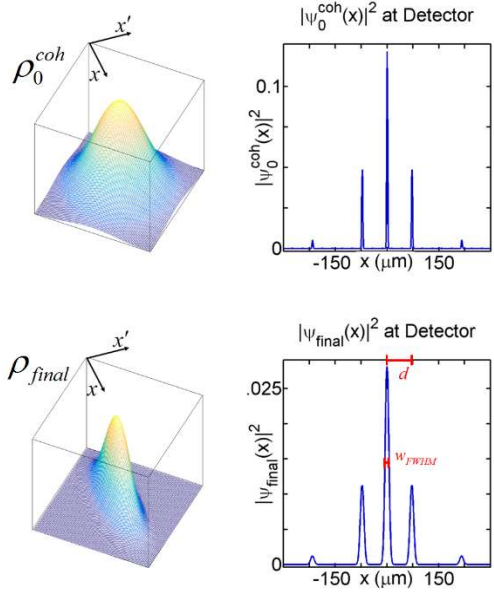

FIG. 4. Coherence reduction. Top Left: Density matrix of a coherent Gaussian electron beam. Top Right: Grating diffraction pattern in the far field after Fourier transformation of the coherent state. Bottom Left: Final density matrix after decoherence evolution according to Equation 2. Bottom Right: Grating diffraction pattern in the far field after Fourier transformation of the deconvoluted partial coherent state.

## VISUALIZATION OF LOSS OF COHERENCE

In order to highlight the loss of contrast in the diffraction pattern, the accumulated image of the MCP detector that was taken by the CCD camera was transformed into the revised images shown in Fig. 2 and Fig. 3 in the main paper [1]. Figure 5 shows the images before and after this process. Line-outs of the image are extracted to obtain diffraction patterns. The line-outs are taken at a slant with the $x$-direction to compensate for image skew. This skew can be explained by small rotational misalignments between the optical elements in the system, however this does not affect the measured coherence length. In the $y$-direction a 4.8 μm range on the detector is integrated for each line-out. Each of these line-outs then correspond to an individual horizontal line on the diffractogram.

After the individual line-outs are fitted according to equation 2 in the main paper, the background term is subtracted from the line-out to show only the relative broadening. Each diffraction peak is normalized by its maximum intensity value for that order.

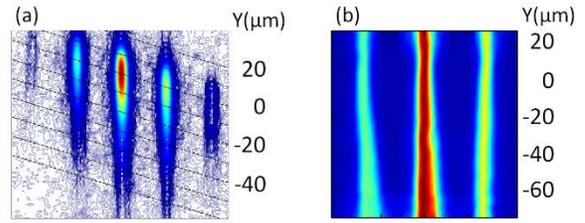

FIG. 5. Visualization of the loss of contrast. (a) Contour of data accumulated by CCD camera. (b) Resulting diffractogram based on data.

---